\documentclass[aps, prd, showpacs, twocolumn]{revtex4-1}

\usepackage{amsmath}
\usepackage{amsfonts}
\usepackage{amssymb}
\usepackage{xcolor}

 \newcommand{\bq}{\begin{equation}}
 \newcommand{\eq}{\end{equation}}
 \newcommand{\bqn}{\begin{eqnarray}}
 \newcommand{\eqn}{\end{eqnarray}}
 \newcommand{\nb}{\nonumber}

\begin{document}

%\preprint{}

\title{Spectral dimension of bosonic string theory}


\author{D. G. Moore}
\email{Douglas{\textunderscore}Moore1@baylor.edu}
%\thanks{}
\affiliation{Department of Physics, Baylor University, Waco, TX, 76798-7316, USA.}

\author{V. H. Satheeshkumar}
\email{VH{\textunderscore}Satheeshkumar@baylor.edu}
%\thanks{}
\affiliation{Department of Physics, Baylor University, Waco, TX, 76798-7316, USA.}


\date{\today}

\begin{abstract}
Given that the scale of quantum gravity is not experimentally accessible, one naturally resorts to mathematical consistency as a measure for a good candidate theory to replace General Relativity at high energies. Reproducing the semi-classical results of black hole entropy has become a standard test for any prospective theory of quantum gravity. It is often argued that another such commonality, albeit less known, is the similar fractal behaviour. It is shown that many, if not all, approaches to quantum gravity predict a spectral dimension of 2 in ultraviolet regime. In this paper, by computing the heat kernel, we show that the spectral dimension of closed bosonic string theory is 26. We discuss the implications of this disparity.
\end{abstract}

%\keywords{gravitational aspects of String theory, quantum gravity, Fractals,  	Field theories in dimensions other than four }

\maketitle

\section{Introduction}
\renewcommand{\theequation}{1.\arabic{equation}} \setcounter{equation}{0}
%%%%%%%%%%%%%%%%%%%%%%%%%%%%%%%%%%%%%%%%%%%%%%%%%%%%%%%%%%%%%%%%%%%%%%%%%
%%%%%%%%%%%%%%%%%%%%%%%%%%%%%%%%%%%%%%%%%%%%%%%%%%%%%%%%%%%%%%%%%%%%%%%%%
%%%%%%%%%%%%%%%%%%%%%%%%%%%%%%%%%%%%%%%%%%%%%%%%%%%%%%%%%%%%%%%%%%%%%%%%%
This year marks the thirtieth anniversary of the Green-Schwarz paper \cite{Green:1984sg} on anomaly cancellation which convinced many theoretical physicists that string theory was a very promising candidate for unifying all the fundamental interactions in nature and prompted an era of intense research in the field. This is often called the ``First Superstring Revolution.'' String theory provides an ultraviolet finite and order-by-order perturbative renormalizable theory of gravity. The problem of non-renormalizability resulting from the direct quantization of General Relativity is well known. Circumventing this disastrous difficulty without losing mathematical consistency is what makes String Theory arguably the most interesting candidate for Quantum Gravity.

The dynamics of the string is described by a two-dimensional worldsheet in the D-dimensional spacetime.  The worldsheet is central to all the physics of the string. When a closed string moves in a curved spacetime its coordinates feel the curvature. In order for there to be a consistent quantum theory, the target spacetime must be a solution to the Einstein field equations. Besides requiring General Relativity to be a part of the theory, it adds corrections to it. General covariance of spacetime becomes an emergent concept in String Theory.

The seemingly different approaches to quantum gravity have a few things in common, including the defining spin-2 graviton and Hawking-Bekenstein entropy of black holes. In addition, it has recently been found that Causal Dynamical Triangulations \cite{Ambjorn:2005db}, Asymptotic Safety \cite{Lauscher:2005qz},  Loop Quantum Gravity \cite{Modesto:2008jz}, Ho\v{r}ava-Lifshitz Theory \cite{Horava:2009if} and also Liouville Quantum Gravity \cite{RhodesVargas} predict the same spectral dimension of 2 \cite{Carlip:2011uc}. It has been suggested that such an agreement must contain some hints of a full theory of quantum gravity \cite{Carlip:2012md}. One exceptional theory is Non-commutative Geometry \cite{Benedetti:2008gu} which predicts spectral dimension of 3 in the ultraviolet regime.

In this context, the often-quoted result from string theory is that of Atick and Witten \cite{Atick:1988si}. They studied the statistical mechanics of string theory: bosonic, type II and heterotic. Such results in the bosonic case were first obtained by Sathiapalan \cite{Sathiapalan:1986db} and Kogan \cite{Kogan:1987jd}. Here we only concentrate on the bosonic case. Their motivation for studying the thermal ensemble is to understand the underlying degrees of freedom in string theory. To this end, they compute the free energy $F = - T \ln{Z}$, where $T$ is the temperature in natural units and $Z$ is the partition function. In a free field theory in D-dimensions, for large $T$, the free energy per unit volume has the form,
\bqn
\frac{FT}{V} \sim T^{D-1}.
\eqn
The interactions will not make this any significantly lesser. In fact, below Hagaedorn temperature, the free energy grows much faster with temperature because of proliferation of string modes. However, above Hagaedorn temperature, the free energy grows linearly with temperature, i.e.,
\bqn
\frac{FT}{V} \sim T.
\eqn 
This implies that the theory undergoes a phase transition only to act like a (1+1)-dimensional quantum field theory at each point of a lattice. That is to say, the high-temperature limit of the free energy in string theory is much less than in any known relativistic quantum field theory. But the effective string theory governing the high-temperature behaviour is still 26 dimensional. This last point is what is not mentioned whenever this result is quoted in the context of spectral dimension. 

 In order to reconcile with other approaches, we compute the spectral dimension in the similar way as other approaches to quantum gravity. Besides throwing light on the spectral behaviour of string theory, our calculations done using heat kernels precisely emphasize the last point of Atick and Witten.

%%%%%%%%%%%%%%%%%%%%%%%%%%%%%%%%%%%%%%%%%%%%%%%%%%%%%%%%%%%%%%%%%%%%%%%%%
%%%%%%%%%%%%%%%%%%%%%%%%%%%%%%%%%%%%%%%%%%%%%%%%%%%%%%%%%%%%%%%%%%%%%%%%%
%%%%%%%%%%%%%%%%%%%%%%%%%%%%%%%%%%%%%%%%%%%%%%%%%%%%%%%%%%%%%%%%%%%%%%%%%
\section{Heat Kernel and Spectral Dimension}
\renewcommand{\theequation}{2.\arabic{equation}} \setcounter{equation}{0}
%%%%%%%%%%%%%%%%%%%%%%%%%%%%%%%%%%%%%%%%%%%%%%%%%%%%%%%%%%%%%%%%%%%%%%%%%
%%%%%%%%%%%%%%%%%%%%%%%%%%%%%%%%%%%%%%%%%%%%%%%%%%%%%%%%%%%%%%%%%%%%%%%%%
%%%%%%%%%%%%%%%%%%%%%%%%%%%%%%%%%%%%%%%%%%%%%%%%%%%%%%%%%%%%%%%%%%%%%%%%%
Besides the obvious applications in many branches of engineering, the study of heat kernel is of importance for algebraic topologists and differential geometers on the side of mathematics; and for quantum field theorists and general relativists on the side of physics. {The study of the spectral theory of the Laplacian through the heat equation was, for many, popularized by Marc Kac \cite{MarcKac}}.

%Following the discussion in Ref.{\cite{Kirsten}}, 
A diffusion process on a $D$-dimensional smooth manifold $\cal{M}$ with boundary $\cal{\partial M}$ and metric $g_{\mu\nu}$ is described by a heat equation
\begin{equation}
\left(\frac{\partial\ }{\partial s} - \Delta\right) K(x,x';s) =0 \quad
\hbox{with $K(x,x',0) = \delta(x-x')$} ,
\label{a1}
\end{equation}
where the Laplace-Beltrami operator is given as, 
\bqn
\Delta = \frac{1}{\sqrt{|g|}} \partial_\mu \left( \sqrt{|g|} g^{\mu \nu} \partial_\nu \right).
\eqn
The function $K(x,x';s)$ is called the heat kernel, which satisfies certain boundary conditions and describes the probability for a random walker to go from point $x$ to $x'$ on the manifold in time $s$. 

One of the basic defining properties of a manifold is its Hausdorff dimension, sometimes simply called the dimension. The spectral dimension is defined as the effective dimension of a diffusion process. In other words, it is the dimensions perceived by a random walker, a randomly moving particle. Mathematically it is defined as
\bqn
\label{SpctDim}
d_s = -2 \lim_{s \rightarrow 0} \frac{d\ln K(x, x';s)}{ d\ln s}.
\eqn

On a smooth manifold the spectral dimension is the same as Hausdorff dimension, but they are generally different on a fractal \cite{Dunne}. There are many ways of computing a heat kernel \cite{Kirsten} and one of the most popular methods in quantum field theory is using the Feynman path integral \cite{FeynmanHibbs}.

As an example, we derive the spectral dimension of a D-dimensional Euclidean space. On a flat manifold $\cal{M} = \mathbb{R}^D$, the heat kernel for a scalar field of mass $m$ \cite{Vassilevich:2003xt} is given by,
\bqn
\label{EclKernel}
K(x,x';s) = (4 \pi s)^{-D/2} \exp{\left(- \frac{(x-x')^2}{4s} - m^2 s \right)} \nb \\
\eqn
where $(x-x')^2$ is essentially the square of the geodesic distance, and the associated Laplacian is 
\bqn
\Delta = - g_{\mu \nu} \nabla^\mu \nabla^\nu + m^2.
\eqn
In order to calculate $d_s$, we first take the logarithm of the kernel Eq.(\ref{EclKernel}), i.e., 
\bqn
d_s = -2 \lim_{s \rightarrow 0} \frac{d}{ d\ln s}\left[ -\frac{D}{2} \ln{4\pi} -\frac{D}{2} \ln{s} - \frac{(x-x')^2}{4s} - m^2 s \right]\nb
\eqn
Now differentiating with respect to $\ln{s}$ and taking the limit as $s \rightarrow 0$ and also requiring the geodesic distance $(x-x') \rightarrow 0$, we get
\bqn
d_s = D \nb
\eqn
which is same as its Hausdorff dimension as expected.

%%%%%%%%%%%%%%%%%%%%%%%%%%%%%%%%%%%%%%%%%%%%%%%%%%%%%%%%%%%%%%%%%%%%%%%%%
%%%%%%%%%%%%%%%%%%%%%%%%%%%%%%%%%%%%%%%%%%%%%%%%%%%%%%%%%%%%%%%%%%%%%%%%%
%%%%%%%%%%%%%%%%%%%%%%%%%%%%%%%%%%%%%%%%%%%%%%%%%%%%%%%%%%%%%%%%%%%%%%%%%
\section{Spectral Dimension of Closed Bosonic String Theory}
\renewcommand{\theequation}{3.\arabic{equation}} \setcounter{equation}{0}
%%%%%%%%%%%%%%%%%%%%%%%%%%%%%%%%%%%%%%%%%%%%%%%%%%%%%%%%%%%%%%%%%%%%%%%%%
%%%%%%%%%%%%%%%%%%%%%%%%%%%%%%%%%%%%%%%%%%%%%%%%%%%%%%%%%%%%%%%%%%%%%%%%%
%%%%%%%%%%%%%%%%%%%%%%%%%%%%%%%%%%%%%%%%%%%%%%%%%%%%%%%%%%%%%%%%%%%%%%%%%
We consider the simplest and most basic class of string theories: closed bosonic string theory in Minkowski background. The graviton appears as a particular state of the closed string. It is described by the Polyakov action
\bqn
\label{PolyakovAction}
S[g, X] = \frac{1}{2} \int{d^2\sigma} \sqrt{g} \, g^{\alpha \beta} \partial_\alpha X^\mu(\sigma) \partial_\beta  X^\nu(\sigma) \eta_{\mu \nu}(X),\nb \\
\eqn
where $g_{\alpha \beta}$ is the worldsheet metric with determinate $g$, $X^\mu$ are local coordinates on target spacetime, $\sigma^\alpha$ are the worldsheet coordinates, and $\eta_{\mu \nu}$ is the D-dimensional Minkowski metric. It is invariant under worldsheet diffeomorphism, Weyl rescalings of the metric and Poincar\'{e} ́ transformations of the target space.

Invariance under Weyl transformations implies that the two-dimensional classical field theory
described by the action in Eq. (\ref{PolyakovAction}) is a conformal field theory. However, demanding conformal invariance after quantization leads to severe constraints on the theory and makes the theory only consistent in $D=26$ dimensions. The worldsheet action can be thought of as a field theory in two dimensions with $26$  scalar fields, while the target spacetime metric behaves like `coupling constants' \cite{Horowitz:2004rn}.

%%%%%%%%%%%%%%%%%%%%%%%%%%%%%%%%%
%%%%%%%%%%%%%%%%%%%%%%%%%%%%%%%%%
\subsection{Heat equation method}
%%%%%%%%%%%%%%%%%%%%%%%%%%%%%%%%%
%%%%%%%%%%%%%%%%%%%%%%%%%%%%%%%%%

In order to find the spectral dimension, we need the heat kernel. One can follow the usual method and obtain the kernel by solving the heat equation constructed with a suitable Laplace-Beltrami operator.  Following this method, we derive a heat equation given by,
\bqn
%\frac{\partial}{\partial\lambda} K[X, \lambda] = \frac{1}{2} \sum_{m}{\left[ \frac{\partial^2}{\partial X_m \partial X_{-m}} - (2 \pi m)^2 X_m X_{-m} \right] K[X, \lambda]}.
\frac{\partial}{\partial\lambda} K[X, \lambda] = \frac{1}{2} \left(\eta^{\mu\nu}\frac{\delta^2}{\delta X^\mu \delta X^\nu} - \eta_{\mu\nu}\frac{\partial X^\mu}{\partial \sigma}\frac{\partial X^\nu}{\partial \sigma}\right) K[X, \lambda]. \nb \\
\eqn
This agrees for the ghost-free case with the results, obtained in two different ways, in the references \cite{Carlip:1988xk} and \cite{Trisnadi:1989tp}. The solution of this obtained using Fourier transforms is given by,
\bqn
\label{Kernel1}
K(X,X'; \lambda) = (4 \pi \lambda)^{-13} \exp{\left(- \frac{(X-X')^2}{2\lambda} - \frac{m^2 \lambda}{2} \right)}. \nb \\
\eqn
with $m^2 \equiv \eta_{\mu\nu}\frac{\partial X^\mu}{\partial \sigma}\frac{\partial X^\nu}{\partial \sigma}$. Substituting this kernel into Eq. (\ref{SpctDim}), we get 
\bqn
d_s = 26.
\eqn

%%%%%%%%%%%%%%%%%%%%%%%%%%%%%%%%%
%%%%%%%%%%%%%%%%%%%%%%%%%%%%%%%%%
\subsection{Path integral method}
%%%%%%%%%%%%%%%%%%%%%%%%%%%%%%%%%
%%%%%%%%%%%%%%%%%%%%%%%%%%%%%%%%%
%\textcolor{red}{
%Although nothing sensible comes out of off-shell propagators in QFT, styring theory does make.... \cite{Nima}
%}

 One can employ the method developed by Feynman \cite{FeynmanHibbs} to compute the kernel which satisfies the `imaginary-time' Sch\"{o}rdinger equation.  It is straightforward, although it involves tedious book keeping. 
This way, one generally computes the heat kernel and from which the propagator or amplitude is obtained. We do the reverse here. The ghost-free amplitude for closed bosonic string  was first computed by Cohen, Moore, Nelson and Polchinski \cite{Cohen:1985sm}. See \cite{Ordonez:1987ep} for closed-string propagator with ghosts. The heat kernel extracted from such an amplitude \cite{Nima} is given by,
\bqn
\label{HK1}
&& K[X_i, X_f; \lambda] = \frac{e^{4\pi\lambda}}{\lambda^{13}} \prod_{n=1}^{\infty}{\left[ 1 - e^{-4\pi n \lambda} \right]^{-24}}  \nb\\
&& \cdot \exp{}\left\{ \frac{-1}{4 \pi \alpha' }  \sum_{m=-\infty}^{\infty} \frac{2 \pi m}{\sinh(2\pi m \lambda)} \right. \nb\\
&& \left. \left[ (|X_m^i|^2 + |X_m^f|^2) \cosh(2\pi m \lambda)  -  2 \Re(X_m^i \cdot X_m^{*f}) \right] \right\} \nb \\
\eqn
where $\lambda$ is the moduli parameter which plays the role of imaginary time. 

We introduce the well-known Dedekind $\eta$-function defined in the upper half complex plane $\cal{H}$. For $s  \in \cal{H}$ it is given by
\bqn
\eta(s) &=& e^{\pi i s/12} \prod_{n=1}^{\infty}{\left[ 1 - e^{2 \pi i n s} \right]}
\eqn
Its twenty fourth power is a modular form of weight 12 which is invariant under the action of group SL$(2, \mathbb{Z})$ and lies at the heart of reasoning that lead to critical dimension of 26 in bosonic string theory. We refer the interested reader to the article by Atiyah \cite{Atiyah} for its many interesting properties. Now making a change of variable $ s \rightarrow 2i\lambda$ and raising it to power $-24$, we get
\bqn
\label{Dedekind24}
\eta(2i\lambda)^{-24} &=& e^{4\pi\lambda} \prod_{n=1}^{\infty}{\left[ 1 - e^{-4\pi n \lambda} \right]^{-24}}
 \eqn

In the path integral notation, we identify the argument of the last exponential in Eq.(\ref{HK1}) as,
\bqn
\label{Sp1}
&&S_P[X; \lambda] =  \frac{-1}{4 \pi \alpha' }  \sum_{m=-\infty}^{\infty} \frac{2 \pi m}{\sinh(2\pi m \lambda)} \nb\\
&& ~~~~ \left[ (|X_m^i|^2 + |X_m^f|^2) \cosh(2\pi m \lambda)  -  2 \Re(X_m^i \cdot X_m^{*f}) \right] \nb \\
\eqn

Putting together these pieces, the heat kernel Eq.(\ref{HK1}) takes a simple form
\bqn
\label{HK2}
&& K[X_i, X_f; \lambda] = \lambda^{-13} \cdot \eta(2i\lambda)^{-24}  \cdot \exp{S_P[X; \lambda]}.
\eqn

The only term that contributes to the spectral dimension in this expression is $\lambda^{-13}$ in the pre-factor. Upon taking the logarithm and differentiating with respect to $\ln\lambda$, all but one term vanish, giving the same answer of $d_s = 26.$ See Appendix for the mathematical details. It is somewhat reassuring that Atick and Witten argue that the effective string theory governing the high-temperature behaviour is indeed 26 dimensional.  See \cite{Calcagni:2013eua}, for a different point of view.

Since string theory involves smooth manifolds, it is expected to have the same number of spectral as well as Hausdorff dimensions. However, in the case of superstring theories, it is interesting to investigate if supersymmetry alters this smoothness of the manifold. This study will be presented elsewhere. 

%%%%%%%%%%%%%%%%%%%%%%%%%%%%%%%%%%%%%%%%%%%%%%%%%%%%%%%%%%%%%%%%%%%%%%%%%
%%%%%%%%%%%%%%%%%%%%%%%%%%%%%%%%%%%%%%%%%%%%%%%%%%%%%%%%%%%%%%%%%%%%%%%%%
%%%%%%%%%%%%%%%%%%%%%%%%%%%%%%%%%%%%%%%%%%%%%%%%%%%%%%%%%%%%%%%%%%%%%%%%%
\section{Conclusions}
\renewcommand{\theequation}{4.\arabic{equation}} \setcounter{equation}{0}
%%%%%%%%%%%%%%%%%%%%%%%%%%%%%%%%%%%%%%%%%%%%%%%%%%%%%%%%%%%%%%%%%%%%%%%%%
%%%%%%%%%%%%%%%%%%%%%%%%%%%%%%%%%%%%%%%%%%%%%%%%%%%%%%%%%%%%%%%%%%%%%%%%%
%%%%%%%%%%%%%%%%%%%%%%%%%%%%%%%%%%%%%%%%%%%%%%%%%%%%%%%%%%%%%%%%%%%%%%%%%
Within classical physics, the role of spacetime has changed radically over time. In non-relativistic classical physics, space is an inert background and time is a monotonically increasing variable unaffected by anything and everything. In special relativity, the distinction between space and time disappears. Whereas in general relativity, spacetime is dynamical and plays an utmost central role. Even in quantum physics, the role of spacetime varies from non-relativistic quantum mechanics to relativistic quantum mechanics to quantum field theory. The role of spacetime in string theory is totally different from that of any other theory. String theory has symmetries which equate spacetimes of different dimensions, geometry and topology. The number of dimensions is fixed by mathematical consistency and there is a provision for reducing the number of dimensions too. Bosonic and fermionic modes ``see'' different number of spacetime dimensions.

Why do some quantum gravity theories behave like two dimensional theories at very high energies? Is it a strange coincidence? Or is there some common symmetry among them that is responsible for such a result? Why is it that some other theories of quantum gravity differ from that? Or, could this approach imply the sigma model is inadequate to reveal something novel about the micro structure of spacetime in string theory?  Trying to make sense of these differences is worth studying and inquiries in this direction may help us to understand the nature of quantum gravity at a deeper level.

%%%%%%%%%%%%%%%%%%%%%%%%%%%%%%%%%%%%%%%%%%%%%%%%%%%%%%%%%%%%%%%%%%%%%%%%%
%%%%%%%%%%%%%%%%%%%%%%%%%%%%%%%%%%%%%%%%%%%%%%%%%%%%%%%%%%%%%%%%%%%%%%%%%
\section*{Acknowledgements}
%%%%%%%%%%%%%%%%%%%%%%%%%%%%%%%%%%%%%%%%%%%%%%%%%%%%%%%%%%%%%%%%%%%%%%%%%
%%%%%%%%%%%%%%%%%%%%%%%%%%%%%%%%%%%%%%%%%%%%%%%%%%%%%%%%%%%%%%%%%%%%%%%%%
We are grateful to Gerald Cleaver for careful reading of the draft and  useful comments. Our special thanks are due to Dwight Russell for not minding us spending inordinate amount of time in the astronomy lab where we did most of our work. One of us (V.H.S.) would like to thank Anzhong Wang for drawing his attention to the study of spectral dimensions. He is also benefited by several discussions during PiTP-2014 at the Institute for Advanced Study in Princeton and Strings-2014 at Princeton University. 

%%%%%%%%%%%%%%%%%%%%%%%%%%%%%%%%%%%%%%%%%%%%%%%%%%%%%%%%%%%%%%%%%%%%%%%%%
%%%%%%%%%%%%%%%%%%%%%%%%%%%%%%%%%%%%%%%%%%%%%%%%%%%%%%%%%%%%%%%%%%%%%%%%%
\section*{Appendix: Limits}
\renewcommand{\theequation}{A.\arabic{equation}} \setcounter{equation}{0}
%%%%%%%%%%%%%%%%%%%%%%%%%%%%%%%%%%%%%%%%%%%%%%%%%%%%%%%%%%%%%%%%%%%%%%%%%
%%%%%%%%%%%%%%%%%%%%%%%%%%%%%%%%%%%%%%%%%%%%%%%%%%%%%%%%%%%%%%%%%%%%%%%%%
We evaluate the limits of the last two terms in the heat kernel expression Eq.(\ref{HK2}). Following \cite{Polchinski}, we expand Eq.(\ref{Dedekind24}) as follows
\bqn
\eta(2i\lambda)^{-24} &=& e^{4\pi\lambda} + 24 + O(e^{-4\pi\lambda})
\eqn
Taking natural logarithm of this expression with respect to $\ln\lambda$, we have
\bqn
\frac{d}{d\ln{\lambda}}\ln\left(\eta(2i\lambda)^{-24}\right)  &=&  \lambda [4\pi + O(-4\pi)]
%&=& \lambda \frac{d}{d\lambda}\ln\eta(2i\lambda)^{-24} \nb\\
\eqn
Now taking the limit of this expression, we get 
\bqn
\lim_{\lambda \longrightarrow 0} \frac{d}{d\ln{\lambda}}\ln\left(\eta(2i\lambda)^{-24}\right)  &=& 0
\eqn

The last term is tackled in the same way. After some manipulations using hyperbolic identities, we can express Eq.(\ref{Sp1}) as below,

\bqn
\label{Sp2}
S_P[X; \lambda] &=&  \frac{-1}{4 \pi \alpha' }  \sum_{m=-\infty}^{\infty} \pi m \left[ (|X_m^f| - |X_m^i|)^2 \coth(\pi m \lambda) \right. \nb\\
&& ~~~~~~~~~~~~ \left. + (|X_m^i| + |X_m^f|)^2 \tanh(\pi m \lambda) \right] 
\eqn

%\bqn
%\frac{d \ln{e^{S_P[X_i, X_f; \lambda]}}}{d\ln \lambda}  &=& \frac{d S_P[X_i, X_f; \lambda]}{d\ln \lambda} \nb
%\eqn
\bqn
\frac{d \ln{e^{S_P[X_i, X_f; \lambda]}}}{d\ln \lambda} &=&  \frac{\lambda}{4 \pi \alpha' }  \sum_{m=-\infty}^{\infty} \pi^2 m^2 \left[ \frac{(|X_m^f| - |X_m^i|)^2}{\sinh^2(\pi m \lambda)}   \right. \nb\\
&& ~~~~~~~~~~~~ \left. - \frac{(|X_m^f| + |X_m^i|)^2}{\cosh^2(\pi m \lambda)} \right] 
\eqn
In the limit as $\lambda \longrightarrow 0 $ and $(|X_m^f| - |X_m^i|) \longrightarrow 0 $, we obtain 

\bqn
\lim_{\lambda \longrightarrow 0} \frac{d S_P[X; \lambda]}{d\ln \lambda} \Bigg\vert_{(|X_m^f| - |X_m^i|) \longrightarrow 0} &=&  0
\eqn

%%%%%%%%%%%%%%%%%%%%%%%%%%%%%%%%%%%%%%%%%%%%%%%%%%%%%%%%%%%%%%%%%%%%%%%%%
%%%%%%%%%%%%%%%%%%%%%%%%%%%%%%%%%%%%%%%%%%%%%%%%%%%%%%%%%%%%%%%%%%%%%%%%%
%%%%%%%%%%%%%%%%%%%%%%%%%%%%%%%%%%%%%%%%%%%%%%%%%%%%%%%%%%%%%%%%%%%%%%%%%
%%%%%%%%%%%%%%%%%%%%%%%%%%%%%%%%%%%%%%%%%%%%%%%%%%%%%%%%%%%%%%%%%%%%%%%%%


\begin{thebibliography}{nbound}
%%%%%%%%%%%%%%%%%%%%%%%%%%%%%%%%%%%%%%%%%%%%%%%%%%%%%%%%%%%%%%%%%%%%%%%%%
%%%%%%%%%%%%%%%%%%%%%%%%%%%%%%%%%%%%%%%%%%%%%%%%%%%%%%%%%%%%%%%%%%%%%%%%%
%%%%%%%%%%%%%%%%%%%%%%%%%%%%%%%%%%%%%%%%%%%%%%%%%%%%%%%%%%%%%%%%%%%%%%%%%
%%%%%%%%%%%%%%%%%%%%%%%%%%%%%%%%%%%%%%%%%%%%%%%%%%%%%%%%%%%%%%%%%%%%%%%%%

\bibitem{Green:1984sg} 
  M.~B.~Green and J.~H.~Schwarz,
  %``Anomaly Cancellation in Supersymmetric D=10 Gauge Theory and Superstring Theory,''
  Phys.\ Lett.\ B {\bf 149}, 117 (1984).
  %%CITATION = PHLTA,B149,117;%%
  %2211 citations counted in INSPIRE as of 30 Mar 2014
  
\bibitem{Ambjorn:2005db} 
  J.~Ambjorn, J.~Jurkiewicz and R.~Loll,
  %``The Spectral Dimension of the Universe is Scale Dependent,''
  Phys.\ Rev.\ Lett.\  {\bf 95}, 171301 (2005)
  [hep-th/0505113].
  %%CITATION = HEP-TH/0505113;%%
  %212 citations counted in INSPIRE as of 26 Feb 2014  
  
\bibitem{Lauscher:2005qz} 
  O.~Lauscher and M.~Reuter,
  %``Fractal spacetime structure in asymptotically safe gravity,''
  JHEP {\bf 0510}, 050 (2005)
  [hep-th/0508202].
  %%CITATION = HEP-TH/0508202;%%
  %151 citations counted in INSPIRE as of 26 Feb 2014  
  
\bibitem{Modesto:2008jz} 
  L.~Modesto,
  %``Fractal Structure from the Area Spectrum,''
  Class.\ Quant.\ Grav.\  {\bf 26}, 242002 (2009)
  [arXiv:0812.2214 [gr-qc]].
  %%CITATION = ARXIV:0812.2214;%%
  %56 citations counted in INSPIRE as of 26 Feb 2014  
  
\bibitem{Horava:2009if} 
  P.~Horava,
  %``Spectral Dimension of the Universe in Quantum Gravity at a Lifshitz Point,''
  Phys.\ Rev.\ Lett.\  {\bf 102}, 161301 (2009)
  [arXiv:0902.3657 [hep-th]].
  %%CITATION = ARXIV:0902.3657;%%
  %328 citations counted in INSPIRE as of 26 Feb 2014  
  
\bibitem{RhodesVargas} 
  R.~Rhodes and V.~Vargas,   
  %``Spectral Dimension of Liouville Quantum Gravity,''
  {Annales Henri Poincar\'{e}} {1424-0637}, {1-18} {(2013)} [arXiv:1305.0154[math]]
  %doi={10.1007/s00023-013-0308-y},
  %url={http://dx.doi.org/10.1007/s00023-013-0308-y},
  %publisher={Springer Basel},}
  
\bibitem{Carlip:2011uc} 
  S.~Carlip and D.~Grumiller,
  %``Lower bound on the spectral dimension near a black hole,''
  Phys.\ Rev.\ D {\bf 84}, 084029 (2011)
  [arXiv:1108.4686 [gr-qc]].
  %%CITATION = ARXIV:1108.4686;%%
  %8 citations counted in INSPIRE as of 26 Feb 2014

\bibitem{Carlip:2012md} 
  S.~Carlip,
  %``Spontaneous Dimensional Reduction?,''
  AIP Conf.\ Proc.\  {\bf 1483}, 63 (2012)
  [arXiv:1207.4503 [gr-qc]].
  %%CITATION = ARXIV:1207.4503;%%
  %4 citations counted in INSPIRE as of 26 Feb 2014

\bibitem{Benedetti:2008gu} 
  D.~Benedetti,
  %``Fractal properties of quantum spacetime,''
  Phys.\ Rev.\ Lett.\  {\bf 102}, 111303 (2009)
  [arXiv:0811.1396 [hep-th]].
  %%CITATION = ARXIV:0811.1396;%%
  %56 citations counted in INSPIRE as of 26 Mar 2014

%%%%%%%%%%%%%%%%%%%%%%%%%%%%%%%%%%%%%%%%%%%%%%%%%%%%%%%%%%%%%%%%%%%%%%%%%
%%%%%%%%%%%%%%%%%%%%%%%%%%%%%%%%%%%%%%%%%%%%%%%%%%%%%%%%%%%%%%%%%%%%%%%%%

\bibitem{Atick:1988si} 
  J.~J.~Atick and E.~Witten,
  %``The Hagedorn Transition and the Number of Degrees of Freedom of String Theory,''
  Nucl.\ Phys.\ B {\bf 310}, 291 (1988).
  %%CITATION = NUPHA,B310,291;%%
  %543 citations counted in INSPIRE as of 26 Feb 2014

\bibitem{Sathiapalan:1986db} 
  B.~Sathiapalan,
  %``Vortices on the String World Sheet and Constraints on Toral Compactification,''
  Phys.\ Rev.\ D {\bf 35}, 3277 (1987).
  %%CITATION = PHRVA,D35,3277;%%
  %167 citations counted in INSPIRE as of 07 Jul 2014

\bibitem{Kogan:1987jd} 
  Y.~.I.~Kogan,
  %``Vortices on the World Sheet of a String: Critical Dynamics,''
  JETP Lett.\  {\bf 45}, 709 (1987)
  [Pisma Zh.\ Eksp.\ Teor.\ Fiz.\  {\bf 45}, 556 (1987)].
  %%CITATION = JTPLA,45,709;%%
  %156 citations counted in INSPIRE as of 09 Jul 2014

%%%%%%%%%%%%%%%%%%%%%%%%%%%%%%%%%%%%%%%%%%%%%%%%%%%%%%%%%%%%%%%%%%%%%%%%%
%%%%%%%%%%%%%%%%%%%%%%%%%%%%%%%%%%%%%%%%%%%%%%%%%%%%%%%%%%%%%%%%%%%%%%%%%

\bibitem{MarcKac}
  M.~Kac, 
  %``Can One Hear the Shape of a Drum?'' 
  The American Mathematical Monthly
  Vol. 73, No. 4, Part 2: pp. 1-23 (1966).

\bibitem{Dunne}
  G.~V.~Dunne,
  %``Heat kernels and zeta functions on fractals,''  
  J. Phys. A: Math. Theor. {\bf 45}, 374016 (2012) 
  %doi:10.1088/1751-8113/45/37/374016
  
\bibitem{Kirsten}
  K.~Kirsten, 
  {\it Spectral functions in mathematics and physics}, 
  Chapman \& Hall/CRC, Boca Raton, FL, (2002)  

\bibitem{FeynmanHibbs}
  R.~P.~Feynman and A.~R.~Hibbs, 
  {\it Quantum Mechanics and Path Integrals}, 
  Dover Publications; Emended Editon by D.~F.~Styer (2010).

\bibitem{Vassilevich:2003xt} 
  D.~V.~Vassilevich,
  %``Heat kernel expansion: User's manual,''
  Phys.\ Rept.\  {\bf 388}, 279 (2003)
  [hep-th/0306138].
  %%CITATION = HEP-TH/0306138;%%
  %314 citations counted in INSPIRE as of 01 Apr 2014

%%%%%%%%%%%%%%%%%%%%%%%%%%%%%%%%%%%%%%%%%%%%%%%%%%%%%%%%%%%%%%%%%%%%%%%%%
%%%%%%%%%%%%%%%%%%%%%%%%%%%%%%%%%%%%%%%%%%%%%%%%%%%%%%%%%%%%%%%%%%%%%%%%%

\bibitem{Horowitz:2004rn} 
  G.~T.~Horowitz,
  %``Spacetime in string theory,''
  New J.\ Phys.\  {\bf 7}, 201 (2005)
  [gr-qc/0410049].
  %%CITATION = GR-QC/0410049;%%
  %35 citations counted in INSPIRE as of 31 Mar 2014

\bibitem{Carlip:1988xk} 
  S.~Carlip,
  %``Sewing Closed String Amplitudes,''
  Phys.\ Lett.\ B {\bf 209}, 464 (1988).
  %%CITATION = PHLTA,B209,464;%%
  %19 citations counted in INSPIRE as of 31 Mar 2014
  
\bibitem{Trisnadi:1989tp} 
  J.~Trisnadi,
  %``Heat equation for the free closed bosonic string,''
  Phys.\ Rev.\ D {\bf 40}, 4186 (1989).
  %%CITATION = PHRVA,D40,4186;%%

\bibitem{Cohen:1985sm} 
  A.~G.~Cohen, G.~W.~Moore, P.~C.~Nelson and J.~Polchinski,
  %``An Off-Shell Propagator for String Theory,''
  Nucl.\ Phys.\ B {\bf 267}, 143 (1986).
  %%CITATION = NUPHA,B267,143;%%
  %164 citations counted in INSPIRE as of 01 Apr 2014
  
\bibitem{Ordonez:1987ep} 
  C.~R.~Ordonez, M.~A.~Rubin and R.~Zucchini,
  %``Polyakov Path Integrals With Ghosts: Closed Strings and One Loop Amplitudes,''
  Phys.\ Lett.\ B {\bf 215}, 103 (1988).
  %%CITATION = PHLTA,B215,103;%%
  %8 citations counted in INSPIRE as of 01 Apr 2014

\bibitem{Nima} 
  This amplitude holds generally. But as a special case, this can be interpreted as off-shell string propagator. Such a propagator is notoriously singular which is not the case in on-shell amplitudes. So it is not sensible to use this to probe the short distance behaviour of strings. Thanks to Nima Arkani-Hamed for explaining this. 

\bibitem{Atiyah}
   M.~Atiyah, 
   %``The logarithm of the dedekind $\eta$-function,"  
   Math. Ann. 278 (1987): 335-380.
   
\bibitem{Calcagni:2013eua}
  G.~Calcagni and L.~Modesto,
  %``Nonlocality in string theory,''
  arXiv:1310.4957 [hep-th].
  %%CITATION = ARXIV:1310.4957;%%
  %5 citations counted in INSPIRE as of 25 Jul 2014    

\bibitem{Polchinski} 
  \textit{See Eq.(7.4.4) in} J.~Polchinski, \textit{String Theory, Volume I,} Cambridge University Press; first   edition (1998).   

\end{thebibliography}
\end{document}